\documentclass[a4paper,10pt]{iopart}
\usepackage[utf8]{inputenc}
\usepackage{graphicx}
\usepackage{fullpage}
\usepackage{hyperref}
\usepackage{soul}
\usepackage[backend=biber]{biblatex}
\newbibmacro{string+doi}[1]{%
  \iffieldundef{doi}{#1}{\href{http://dx.doi.org/\thefield{doi}}{#1}}}
\DeclareFieldFormat{title}{\usebibmacro{string+doi}{\mkbibemph{#1}}}
\DeclareFieldFormat[article]{title}{\usebibmacro{string+doi}{\mkbibquote{#1}}}
\renewbibmacro{in:}{}

\usepackage[breakable]{tcolorbox}

\newcommand{\rhof}{\bar\rho}
\newcommand{\Dnucs}{\Delta\nu_\textrm{\tiny Cs}}

\begin{document}

\title{Properties of a definition of the SI second based on several optical transitions}
\author{Jérôme Lodewyck}
\address{Laboratoire Temps Espace (LNE-OP), Observatoire de Paris, Université PSL, Sorbonne Université, Université de Lille, LNE, CNRS, 61 avenue de l’Observatoire, 75014 Paris, France}
\ead{jerome.lodewyck@obspm.fr}
\author{Tetsuya Ido}
\address{National Institute of Information and Communications Technology, 4-2-1 Nukui-kitamachi, Koganei, Tokyo, 184-8795 Japan}
\ead{ido@nict.go.jp}

\begin{abstract}
	The current definition of the SI second based on the Cs ground state hyperfine transition is expected to be replaced by a new definition based on optical frequency standards in the next decade. Several options are currently under consideration for the new definition, including a definition based on the weighted geometric mean of several transitions. In this paper, we review several properties of this option, and most notably introduce a quantitative method to set the weights of transitions, and a graphical representation of the unit that make the distinction between the definition and the realisation of the unit easier to understand.
\end{abstract}

\section{Introduction}

The second is one of the seven base unit of International System of units (SI). It is currently defined by fixing the numerical value of the Cs hyperfine transition:

\begin{quotation}
	The second, symbol s, is the SI unit of time. It is defined by taking the fixed numerical value of the caesium frequency $\Dnucs$, the unperturbed ground-state hyperfine transition frequency of the caesium-133 atom, to be 9\,192\,631\,770 when expressed in the unit Hz, which is equal to s$^{-1}$.
\end{quotation}

Since more than a decade, the best realisation of the SI second have been overcome by optical frequency standards, and nowadays, several of these optical frequency standards reach uncertainties up to two orders of magnitude better than caesium fountain clocks~\cite{ludlow2015optical}. A large number of frequency ratios have been measured, involving optical-to-microwave as well as full optical-to-optical ratios, which are exploited to produce recommended frequencies for many optical transitions -- the Secondary Representations of the Second (SRS) -- using a global least square fit~\cite{pub.1058980983} or a graph theory framework~\cite{Robertsson_2016}. Secondary Frequency Standards (SFS) realising these transitions are now extensively used to steer the International Atomic Time (TAI)~\cite{bilicki2017contributing, hachisu2018months}. However, the definition of the second being still based on Cs, the uncertainty of the recommended frequencies of SRSs are limited by the accuracy of Cs standards, at $10^{-16}$, although some optical-to-optical frequency ratios can be measured with an uncertainty in the mid-$10^{-18}$~\cite{boulder2021frequency}. In this context, it is now desirable to adopt a new definition of the SI second, based on optical transitions. To this aim, working groups of the CCTF (Consultative Committee for Time and Frequency) have crafted a roadmap~\cite{10.1088/1681-7575/ad17d2} including criteria to be fulfilled before the SI second can be redefined, as well as a presentation of the different options under consideration for the new definition. The first possibility, named hereunder ``option 1'', simply consists in replacing the caesium clock transition by an optical transition in an atom or an ion, which is yet to be chosen. The second possibility, ``option 2'', is based on the proposition published in~\cite{lodewyck2019definition}, and consists in replacing the caesium clock transition by the weighted geometric mean of a set of clock transitions. The definition would then read:
\begin{quotation}
	The second, symbol s, is the SI unit of time. It is defined by taking the fixed numerical value of the geometric mean of the frequencies of the clocks transitions $\nu_1$, $\nu_2$, \ldots $\nu_n$ with weights $w_1$, $w_2$, \ldots, $w_n$, to be $N$ when expressed in the unit Hz, which is equal to s$^{-1}$.
\end{quotation}
where $N$, the subscripts $1$ to $n$, and $w_i$ are placeholders for the normalisation constant, transitions and weights actually chosen at the time of the redefinition. Formally, this definition is written:
\begin{equation}
    \prod_{k \in \mathcal{C}} \nu_k^{w_k} \equiv N~\textrm{Hz} \qquad \textrm{exactly},
\end{equation}
where $\mathcal{C}$ is the ensemble of transitions composing the unit.

Given the fast progress in optical time and frequency metrology over the last two decades, and the diversity of clock transitions under study, it is possible that the new definition, fixed within a few years, may soon become obsolete, as optical frequency standards are improved. To face this situation, a flexible definition of the second can be considered, in which the species involved in the definition can be regularly adjusted, taking into account the reported uncertainty of measured frequency ratios. By allowing to continuously tune the balance between species, option 2 makes the changes of such a flexible definition smooth. Furthermore, a quantitative criterion can be computed to determine whether the clock uncertainties have progressed enough for an adjustment to be relevant~\cite{lodewyck2019definition}. The wording of such a flexible definition would be:
\begin{quotation}
	The second, symbol s, is the SI unit of time. It is defined by taking the fixed numerical value of the weighted geometric mean of the frequencies of a set of atomic clocks transitions, to be a constant $N$ when expressed in the unit~Hz, which is equal to~s$^{-1}$. The weights and the constant $N$ are published by the International Committee for Weights and Measures (CIPM) and updated according to frequency ratio measurements, in order that the unit converges.
\end{quotation}
Following the denomination used in the roadmap~\cite{10.1088/1681-7575/ad17d2}, this dynamical implementation of option 2 is referred hereunder as option 2b, while the static version is referred to as option 2a.

Option 2 can be viewed as a generalisation of option 1, the former reducing to the latter in the case of a single transition having a unit weight. Option 2 is particularly adapted to the current situation where many optical frequency standards are evenly matched, and highly connected through measured frequency ratios and a global fit thereof. However, the main drawbacks of the option 2 so far are twofold: First it conceptually departs from the traditional concept of fixing the value of a physical signal to set a unit, by rather fixing an average of frequencies, which gives the impression that frequency ratios and their uncertainties enter the definition. Second, the initial proposition~\cite{lodewyck2019definition} does not prescribe a quantitative method to determine the weights of the transitions entering the unit, leaving place for subjective or political debate.

This papers aims at addressing these two shortcomings. In section~\ref{sec:graphical} we propose a graphical representation of option 2  that illustrates how the definition based on several transitions is self-consistent, without any intrinsic uncertainty, while frequency ratios and their uncertainties only enter when linking realisations with a single (or a few) transitions to the definition. In section~\ref{sec:weights}, we propose a quantitative procedure based on the least-squares method, in order to attribute weights to the transitions composing the unit. This procedure is then illustrated by a simulation of option 2 using the 2021 adjustment of global frequency ratios by the CIPM (section~\ref{sec:2021}).

\section{Graphical representation}
\label{sec:graphical}

In this section, we propose a graphical representation of the SI second, which provides an intuitive understanding of the definition, its realisations, and how recommended frequencies are derived from the definition and frequency ratios measurement (section~\ref{sec:graphical_representation}). After the definition is set, new frequency ratio measurements may become available, with improved clock uncertainty. This evolution may lead to updated recommended frequencies, and eventually to a new redefinition, which are graphically explained in section~\ref{sec:graphical_updateratios} and~\ref{sec:graphical_redefinition}, respectively. For each graphical representation, we present options 1 and 2 side by side, in order to highlight their resemblances and their differences.

\subsection{Representation of the definition and frequency ratios in the frequency space}
\label{sec:graphical_representation}

\begin{figure}
	\begin{center}
		\includegraphics[width=0.7\textwidth]{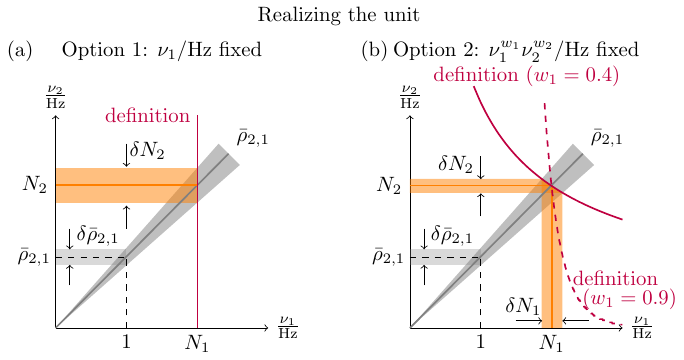}
	\end{center}
	\caption{\label{fig:graphical}
	Graphical representation of options 1 and 2, in the special case of two clock transitions, with frequency $\nu_1$ and $\nu_2$ respectively. In the $(\nu_1/\textrm{Hz}, \nu_1/\textrm{Hz})$ plane, the definition is represented by a purple curve, whose equation is exactly known. The recommended frequencies $N_1$ and $N_2$ are the coordinates of the intersection point between this curve and the most likely frequency ratio $\bar{\rho}_{2,1} \simeq \nu_2/\nu_1$ (slope of the grey line). While the definition itself has not intrinsic uncertainty, the ratio $\rho_{2,1}$ is only known with a limited uncertainty $\delta \bar \rho_{2,1}$ (grey area), resulting in an uncertainty $\delta N_1$ and $\delta N_2$ on the recommended frequencies (orange area). In option 1 (graph (a)), we have $\delta N_1 = 0$ (the realisation of the clock transition 1 is a Primary Frequency Standard), but realising the unit with a Secondary Frequency Standard (clock transition 2 being a Secondary Representation of the Second) incurs an uncertainty $\delta N_2$. In option 2  (graph (b)), the use of the weighted geometric mean of several transitions (here we chose $w_1 = 0.4$ and $w_2 = 0.6$ for the solid purple definition curve), results in a more balanced distribution of this uncertainty among the transitions. With a large weight on $\nu_1$, option 2 falls back close to option 1. This is illustrated by the dashed purple curve for which $w_1 = 0.9$, that would give a thinner uncertainty $\delta N_1$, and a thicker uncertainty $\delta N_2$.}
\end{figure}

At first sight, option 2 significantly differs from option 1, making it less intuitive. However, this difficulty can be alleviated by graphically representing both the definition and the frequency ratios in frequency space, revealing their fundamental differences.
Figure~\ref{fig:graphical} proposes such a graphical representation of options 1 and 2, in the special case of two clock transitions with frequencies $\nu_1$ and $\nu_2$. In these graphs, the definition is represented by a purple curve in the ($\nu_1/$Hz, $\nu_2/$Hz) plane. In option 1 (Figure~\ref{fig:graphical}~(a)), as well as in the current definition of the SI second based on the Cs hyperfine transition, the definition is set by fixing the numerical value in Hz of the frequency of a single transition (here labelled 1) to $\nu_1/$Hz $= N_1$ exactly, and the frequency standards implementing this specific transition are so-called Primary Frequency Standards (PFS). Therefore, the definition curve is a straight vertical line. In option 2, on the other hand, the definition equation is $\left( \nu_1/{\rm Hz}\right)^{w_1} \left( \nu_2/{\rm Hz}\right)^{w_2} = N$ exactly, and as shown in Figure \ref{fig:graphical}~(b), the definition is no more a straight line. Nevertheless, in both options 1 and 2, the definition curve is an exact, conventional curve, that bears no uncertainty, and that is independent of our knowledge of frequency ratios. In fact, choosing a definition that is not represented by a straight line, but rather by a curve can also be encountered if the second were redefined by fixing a fundamental constant, as explained in \ref{sec:graphicaloption3}. Finally, if $w_1$ and $w_2$ tend to 1 and 0, respectively, the definition curve  $\left( \nu_1/{\rm Hz}\right)^{w_1} \left( \nu_2/{\rm Hz}\right)^{w_2} = N$ approaches a vertical straight line as represented by the purple dashed curve in Figure~\ref{fig:graphical}~(b), where $w_1$ and $w_2$ are 0.9 and 0.1, respectively. This dashed curve resembles the definition line of the option 1, illustrating that option 1 is a special case of option 2. Thus, there is no crucial difference between whether the definition line is a vertical straight line or a curve.

In the current definition of the SI second, and expectedly in option 1, the CIPM authorizes several other frequency standards, realising transitions called Secondary Representations of the Second (SRS), to be used as realisations of the second with an uncertainty. To this aim, the CIPM publishes a list of recommended frequencies for the SRSs, determined as follows:

\begin{enumerate}
    \item \label{item:lsq} A least squares analysis~\cite{pub.1058980983} or a graph theory framework~\cite{Robertsson_2016}, taking as input collected world-wide absolute frequencies and frequency ratios measurements, yields a set of most likely frequency ratios $\bar{\rho}_{i,j}$ between all transitions. This analysis also yields uncertainties $\delta\bar \rho_{i,j}$ for these most-likely ratios, representative of their unknown difference with the actual frequency ratios $\rho_{i,j} = \nu_i/\nu_j$. Because this procedure only involves dimension-less quantities, its outcome is independent of the definition of the SI second.
    \item The recommended frequencies of SRSs, whose numerical values in~Hz are noted $N_i$ in this paper, and their uncertainties $\delta N_i$ are deduced from $N_i/N_j = \bar{\rho}_{i,j}$, where $\bar{\rho}_{i,j}$ is the outcome of (\ref{item:lsq}), with the additional constraint given by fixing the Cs transition frequency to 9\,192\,631\,770 Hz exactly.
\end{enumerate}

This process is graphically represented in Figure~\ref{fig:graphical}~(a), where the estimated frequency ratio between the transitions 2 and 1, as determined by (\ref{item:lsq}), is drawn as a black line with slope $\bar{\rho}_{2,1}$, while the uncertainty $\delta \bar \rho_{2,1}$ is graphically represented as a shaded area. Because the ratio of recommended frequencies verifies $N_2/N_1 = \bar{\rho}_{2,1}$, and because option 1 constrains the frequency $\nu_1$ to be $N_1$~Hz by definition, the point $(N_1, N_2)$ is located at the intersection between the purple line (definition) and the black line (frequency ratio). Although the definition equation is exact with no uncertainty, the uncertainty $\delta \bar  \rho_{2,1}$ on the slope $\bar \rho_{2,1}$ causes the point $(N_1, N_2)$ to have a probability distribution, hence and uncertainty. For option 1, this uncertainty is only on $N_2$.

In option 2, the recommended frequencies $N_i$ are similarly determined by the intersection between the curve representing the definition, and the line representing the frequency ratio $\rho_{2,1}$ as determined by the fit procedure (i). This is represented in Figure~\ref{fig:graphical}~(b). Comparing this figure to Figure~\ref{fig:graphical}~(a), it is clear that there is no fundamental difference between options 1 and 2 in the way the recommended frequencies $N_i$ are determined. Also like option 1, the uncertainties $\delta N_1$ and $\delta N_2$ on the recommended frequencies are given by the projections on the horizontal and vertical axis of the intersection between the exact definition curve and the shaded area representing the uncertainty $\delta \bar \rho_{1,2}$. They are quantitatively obtained by taking the logarithmic derivative of $N_2/N_1 = \bar{\rho}_{2,1}$ and of $N_1^{w_1} N_2^{w_2} = N$:
\begin{equation}
    \label{eq:logdervi}
\frac{d N_2}{N_2} - \frac{d N_1}{N_1} = \frac{d \bar{\rho}_{2,1}}{\bar{\rho}_{2,1}},
\quad \textrm{and} \quad w_1 \frac{d N_1}{N_1} + w_2 \frac{d N_2}{N_2} = 0.
\end{equation}
Combing these two equations, we obtain:
\begin{equation}
\label{eq:dN12}
\frac{\delta N_1}{N_1} = w_2 \frac{\delta \bar{\rho}_{2,1}}{\bar{\rho}_{2,1}}, \quad \textrm{and} \quad \frac{\delta N_2}{N_2} = w_1 \frac{\delta \bar{\rho}_{2,1}}{\bar{\rho}_{2,1}}
\end{equation}
Equations~(\ref{eq:dN12}) relate the relative uncertainty of the recommended frequencies to the relative uncertainty on the frequency ratio. For a definition based on an arbitrary number of transitions, the equations~(\ref{eq:dN12}) are generalized by equation~(\ref{eq:dNk}), which gives the uncertainty of the recommended frequencies as a function of the uncertainties of the frequency ratios and their correlations.
Taking $w_1 = 1$ and $w_2 = 0$, we recover the case of option 1, for which $\delta N_1 = 0$ and $\delta N_2 = N_1\delta\bar \rho_{2,1}$, whereas under option 2 both $N_1$ and $N_2$ have an uncertainty. However, as shown by equations~(\ref{eq:dN12}) and as illustrated in Figure~\ref{fig:graphical}, these uncertainties are reduced by the other's transition weight, hence balanced between the two transitions.

\subsection{Update of frequency ratios}
\label{sec:graphical_updateratios}

\begin{figure}
	\begin{center}
		\includegraphics[width=0.7\textwidth]{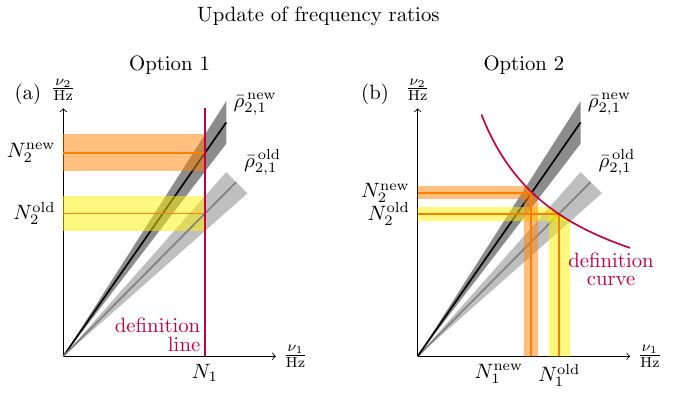}
	\end{center}
    \caption{\label{fig:graphicalupdateratios}
	Update of recommended frequencies $N_i$, triggered by the update of $\bar \rho_{2,1}$ associated with a new least square analysis (in the pictures, the change of frequency ratios is exaggerated for clarity. In reality the change in $\bar \rho_{2,1}$ would probably be small enough for the gray areas to overlap):
	(a): Case of option 1 as well as the current revision of the CIPM recommended frequencies for SRS (here $N_2$). $N_1$ never changes since it is fixed by the definition.
	(b): Case of option 2, where both $N_1$ and $N_2$ change since the update of $\rho_{2,1}$ shifts the point $(N_1, N_2)$ along the fixe definition curve. This graph illustrates that in option 2, as in option 1, the definition in unaffected by a better knowledge of frequency ratio.}
\end{figure}

While the frequency ratio $\rho_{2,1}$ is an invariant dimensionless quantity given by nature, our best estimate $\bar{\rho}_{2,1}$ of this frequency ratio is updated  with new measurements of frequency ratios entering the least squares analysis. Figure \ref{fig:graphicalupdateratios} illustrates how the new estimate of $\rho_{2,1}$ leads to an update of the recommended frequencies, as the intersection point $(N_1, N_2)$ moves along the definition line.

For option 1 (Figure \ref{fig:graphicalupdateratios}(a)) the revision of the estimated frequency ratio $\bar \rho_{2,1}$ results in a vertical shift of $\left( N_1, N_2 \right)$ (the intersection of the slope and the vertical definition line), given that $N_1$ remains constant since $\nu_1$ is the defining transition. This situation currently happens when the CCTF Working Group on Frequency Standards incorporates new frequency ratio measurements in the global least squares fit, and produces updated $\bar \rho_{i,j}$. This leads to a new set of recommended frequencies by the CIPM.

Similarly, for option 2 (Figure \ref{fig:graphicalupdateratios}(b)), $(N_1, N_2)$ moves on the definition curve as the estimated frequency ratio $\bar\rho_{2,1}$ is updated, resulting in updated recommended frequencies. Since this process does not involve a change in the definition curve, it is by no means a redefinition,. It is rather equivalent to the update of CIPM recommended frequencies currently being carried out under the option 1. From $N_2/N_1 = \bar{\rho}_{2,1}$, the change $\Delta \bar \rho_{2,1}$ due to the update of $\bar{\rho}_{2,1}$ can cause corrections in both $N_1$ and $N_2$, denoted as $\Delta N_1$ and $\Delta N_2$, which depend on the weights $w_1$ and $w_2$. Since $(N_1, N_2)$ must always move along the constant definition line $\nu_1^{w_1} \nu_2^{w_2} = N$~Hz, equations~(\ref{eq:logdervi}) yields:
\begin{equation}
    \label{eq:var2}
\frac{\Delta N_1}{N_1} = -w_2 \frac{\Delta \bar{\rho}_{2,1}}{\bar{\rho}_{2,1}}, \qquad \frac{\Delta N_2}{N_2} = w_1 \frac{\Delta \bar{\rho}_{2,1}}{\bar{\rho}_{2,1}}.
\end{equation}

Here, it is reasonable that the fractional changes resulting from the update of $\bar{\rho}_{2,1}$ are distributed over the fractional changes in $N_1$ and $N_2$ according to the weights set by the definition. This is an advantage of the weighted geometric mean, compared to other means such as the arithmetic mean, which require other parameters such as nominal frequencies in order to determine the uncertainty in $N_1$ and $N_2$, as detailed in \ref{sec:artihmvsgeom}. However, even though the definition line changes from a straight line to a curve, the variation in the slope due to experimental update is so small that in option 2, it is practically valid to treat the curve as a linear approximation at the intersection point.

\subsection{Change of definition}
\label{sec:graphical_redefinition}

\newcommand{\new}{{\rm new}}
\newcommand{\old}{{\rm old}}

\begin{figure}
	\begin{center}
		\includegraphics[width=0.7\textwidth]{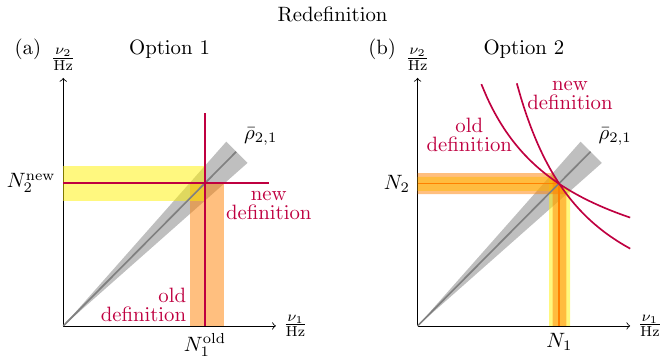}
	\end{center}
    \caption{\label{fig:graphicalredefinition}
	Graphical representation of a redefinition of the unit.
	(a): In option 1, the transition of the definition changes from $\nu_1$ to $\nu_2$ with the change of the associated uncertainty from $\delta N_2$ to $\delta N_1$.
	(b): In option 2, the definition curve is modified by updating the weights or adding or removing some clock transitions to the pool of clock transitions.\\ For both options, the definition curve changes while keeping the recommended frequencies constant, as discussed in section~\ref{sec:graphical_redefinition}: the point $(N_1, N_2)$ is a constant pivot throughout the redefinition.}
\end{figure}

As the uncertainty of clocks improve, a new redefinition may become necessary to account for these improvements. Practically, such a redefinition involves updating the pool of transitions from $\mathcal{C}^\old$ to $\mathcal{C}^\new$, and within these pools, updating the weights of each transition from $w_i^\old$ to $w_i^\new$, (the argument presented here covers both options 2 and 1, the latter begin a special case of the former for which $w_j = 1$ for a specific transition $j$). In doing so, the normalisation constant of the unit must be changed from $N^\old$ to $N^\new$ in order to ensure the best possible continuity of the unit. The relation between $N^\old$ and $N^\new$ is given by~\cite{lodewyck2019definition}:
\begin{equation}
    \label{eq:Nprime}
	N^\new = N^\old \prod_{k \in \mathcal{C^\old}\cup \mathcal{C}^\new} \rhof_{ik}^{w^\old_k - w^\new_k},
\end{equation}
where $\rhof_{ik}$ is the adjusted frequency ratio between transitions $i$ and $k$.
This relation is derived by stating that the recommended frequency of a given transition $i$ is kept constant through the redefinition process:
\begin{equation}
	N_i^\new = N^\new \prod_{k \in \mathcal{C^\new}} \rhof_{ik}^{w^\new_k} = N^\old \prod_{k \in \mathcal{C^\old}} \rhof_{ik}^{w^\old_{k}} = N_i^\old.
\end{equation}
It is noteworthy that the computation of $N^\new$ by equation~(\ref{eq:Nprime}) is independent of the specific transition $i$ chosen as a pivot, and consequently that the numerical values of recommended frequencies of all transitions are invariant when the unit is redefined. This result is actually rather intuitive: the recommended frequency for a given transition represents our best knowledge of the numerical value of the transition frequency in the unit, and the normalisation constant is chosen such that the unit is unchanged 
through the redefinition. In particular, this results holds for an initial and/or final definition based on a single transition: whatever option is chosen for the next redefinition of the second, the recommended frequency for the caesium clock transition right after the redefinition will be 9\,192\,631\,770~Hz; and if a single optical transition is selected, its exact frequency will be equal to its previously recommended frequency, while all other recommended frequencies will be unchanged. Nevertheless, these recommended frequencies, including for caesium, will eventually evolve when more precise frequency ratios are measured.

This redefinition process is illustrated in Figure~\ref{fig:graphicalredefinition} for a set of two transitions. For option 1 (Figure~\ref{fig:graphicalredefinition}~(a)), the change of the defining transition corresponds to switching of the definition line from vertical to horizontal line. In contrast, the change of the definition in option 2 (Figure~\ref{fig:graphicalredefinition}~(b)) involves the addition or removal of transitions, the update of weights $w_i$ for each transition, and the change in the definition constant $N$, resulting in a change of the definition curve. For both options, the new definition curve pivots around $(N_1, N_2)$, as these recommended frequencies are invariant under a redefinition. This is the opposite situation of the effect of updated recommended frequency ratios (section~\ref{sec:graphical_updateratios} and Figure~\ref{fig:graphicalupdateratios}), for which the definition curve is constant but the recommended frequencies change.

However, even though the recommended frequencies are invariant under a redefinition, their uncertainties do change when the unit is redefined. As illustrated in Figure~\ref{fig:graphicalredefinition} where the uncertainties of $N_i$ for the old and new definition are represented in yellow and orange, respectively, this effect arises from the change in the slope of the definition curve. For instance, in option 1 (Figure~\ref{fig:graphicalredefinition}~(a)), the null uncertainty is transferred from $N_1$ to $N_2$, following the change of the defining transition. In option 2 (Figure~\ref{fig:graphicalredefinition}~(a)), all recommended frequencies have a non-zero recommended uncertainty, but their balance changes when the weights are changed (see discussion  in section~\ref{sec:weights})

In the current discussions of the CCTF working groups, two sub-options, option 2a and 2b, are under consideration. Option 2a involves no updates in the selected transitions or weights after the initial redefinition, while option 2b does involve updates. The discussion above suggests that the issue is not whether to choose between the options 2a and 2b, but rather to determine whether or when the change of the definition should occur in the course of updating standard frequencies $N_i$.

In conclusion, option 2 can be understood more clearly through the graphical framework we have introduced in this section. Specifically, we can separate the two processes, namely the updates of $N_i$ and the change of the definition as shown in Figure \ref{fig:graphicalupdateratios}(b) and Figure \ref{fig:graphicalredefinition}(b), respectively. It should also be noted that the figures shown here illustrate only the case of two transitions, and if transitions are added or removed, the dimensionality of the frequency space changes, making it impossible to represent in a two-dimensional graphical picture, although the underlying concept remains the same.

\section{Choice of weights}
\label{sec:weights}

In this section we propose a prescription in order to attribute weights to the various transitions composing the unit. This prescription is needed at the moment of the redefinition, and, in the case of option 2b, as predefined rules for updating the weights when the clock have significantly improved after the initial redefinition.

\subsection{Weights from clock uncertainties}
\label{sec:weightsfromu}

Intuitively, the weight of a transition $i$ should be a decreasing function of the uncertainty $u_i$ of the clocks realising this transition: the most precise clocks should have the largest weight.

When realising the unit with a single clock based on the transition $i$, the total uncertainty is:
\begin{equation}
	\label{eq:utot}
	u_\textrm{\tiny tot}^2 = u_i^2 + \left(\frac{\delta N_i}{N_i}\right)^2,
\end{equation}
that is to say the quadratic sum of the clock uncertainty $u_i$ itself, and the relative uncertainty of the recommended frequency $N_i$. The latter appears as an overhead in the realisation of the unit, coming from the link between the frequency of the transition $i$ and the definition. In option 1, this overhead is zero for the Primary Frequency Standards which realise the transition defining the unit, and non-zero for Secondary Frequency Standards realising SRSs. For SRSs, it is fundamentally limited by the uncertainty of the PFS, because the uncertainty in the realisation of the PFS is transferred to the SRSs in addition to the intrinsic uncertainty of these SRSs via the measurement of frequency ratios. The aim of the unit based on the geometric mean of several transitions we consider here is to suppress this bottleneck by delocalizing the origin of the unit over all transitions. With this definition, the weights should thus be chosen in such a way that the overhead $\delta N_i/N_i$ is always at most comparable to, and ideally negligible with respect to the clock uncertainty $u_i$.

In~\cite{lodewyck2019definition}, it was shown, in the special case of a unit composed of two transitions, that this condition can be satisfied by choosing $w_2/w_1 = (u_1/u_2)^l$ with $l > 1$. From this result, it sounds reasonable to choose weights proportional to the inverse \emph{squared} uncertainty $1/u_i^2$. Here, we show that this choice is actually optimal, in the more general case of an arbitrary number of transitions. For this, we define the residue:
\begin{equation}
	r_i  = \frac{\delta N_i/N_i}{u_i},
\end{equation}
as the ratio between the uncertainty on the link to the definition and the clock uncertainty. The optimal choice of weights is such that these residues are minimal over the whole set of transitions. This optimum is reached by a least square adjustment, minimizing the sum of squared residues using the weights $w_i$ as free parameters:
\begin{equation}
	\label{eq:leastsquaresw}
	\frac{\partial R}{\partial w_i} = 0 \textrm{ for all $i$}, \quad \textrm{with } R = \sum_k r_k^2.
\end{equation}
In~\ref{sec:proofoptimal}, we show that the solution of this set of equations is:
\begin{equation}
	\label{eq:optimalw}
	w_l = \frac{\frac{1}{u_l^2}}{\sum_k \frac{1}{u_k^2}},
\end{equation}
which effectively states that in order to minimize the global overhead in the realisation of the unit, the weight of a transition should be chosen proportional to the inverse of the squared uncertainty of the realisations of this transition.

\subsection{Determination of the clocks' uncertainty}

The two terms of equation~(\ref{eq:utot}) have a very different origin: on the one hand the term $\delta N_i/N_i$, \emph{i.e.} the fractional uncertainty on the recommended frequency for the transition $i$, arises from a global fit of measured frequency ratios, as explained in section~\ref{sec:graphical_representation}, and as detailed in~\ref{sec:globalfit}. On the other hand, $u_i$ is the uncertainty of the local clock realising the unit \emph{via} the implementation of the clock transition $i$. Because this uncertainty is not intrinsic to the transition $i$, but rather varies from realisation to realisation, we now have to specify which value $u_i$ should be used to set the weights of the transition $i$ in equation~(\ref{eq:optimalw}).
In practice, the uncertainty of clocks that have produced frequency ratio measurements have operational and comparison capabilities particularly relevant for the realisation of the SI second, and are central to building the concept of recommended frequencies. Their typical uncertainties $u_i$, that can be used to define the weight $w_i$ according to~(\ref{eq:optimalw}), are therefore contained in the uncertainty of published frequency ratio, and determine the uncertainty of recommended frequency ratios. Here, we propose to derive the typical uncertainties $u_i$ from these uncertainties. For this, we note:
\begin{equation}
	u_{i,j} = \frac{\delta\bar\rho_{i,j}}{\bar\rho_{i,j}}
\end{equation}
the relative uncertainty on the frequency ratio $\rho_{i,j}$. This set of uncertainties is an outcome of the global fit of measured frequency ratios realised by the working groups of the CCTF, as stated by equation~(\ref{eq:deltarhoij}). In~\cite{lodewyck2019definition}, we propose to set $u_i^2$ as the average of the two smallest values of $u_{i,j}^2$ over all possible transitions $j$, based on the heuristic observation that this assumption yields sensible weights. Here, we propose a more rigorous approach, in order to deterministically define the uncertainties $u_i$, hence the weights, from the values $u_{j,k}$. For this, we define the figure of merit:
\begin{equation}
    \label{eq:s}
	s_{j,k} = \frac{u_j^2 + u_k^2}{u_{j,k}^2} - 1.
\end{equation}
For an ideal situation in which all possible frequency ratios have been measured with the best possible clocks, and limited by the systematic uncertainty with no correlations, we expect all these quantities to be zero, as the relative uncertainty $u_{j,k}$ on the frequency ratio $\rho_{j,k}$ is expected to be the quadratic sum of the individual clock uncertainties. For a practical situation, in which only a part of the frequency ratios have been measured, and with clocks with different uncertainties, we propose a least square procedure to find the most suitable values of $u_i$, that minimize the sum of squared residuals $s_{j,k}$:
\begin{equation}
    \label{eq:dS}
	\frac{\partial S}{\partial u_i^2} = 0 \quad \textrm{for all $i$}, \quad \textrm{with } S = \sum_{j> k}s_{j,k}^2.
\end{equation}
These equations can be rewritten as:
\begin{equation}
	\sum_{k > i} \frac{s_{i,k}}{u_{i,k}^2} = 0 \quad \textrm{for all $i$},
\end{equation}
and their solution is:
\begin{equation}
	\label{eq:u2}
	\left[
	\begin{array}{c}
	 u_1^2 \\
	 u_2^2 \\
	 \vdots \\
	 u_n^2
	\end{array}
	\right] = M^{-1}
	\left[
	\begin{array}{c}
	 \xi_1 \\
	 \xi_2 \\
	 \vdots \\
	 \xi_n
	\end{array}
	\right],
	\quad \textrm{where } M =
	\left[
	\begin{array}{cccc}
	 \zeta_1 & u_{1,2}^{-4} & \dots & u_{1,n}^{-4} \\
	 u_{2,1}^{-4} & \zeta_2 & \dots & u_{2,n}^{-4} \\
	 \vdots & \vdots & \ddots & \vdots \\
	 u_{n,1}^{-4} & u_{n,2}^{-4} & \dots & \zeta_n
	\end{array}
	\right],
\end{equation}
and with $\xi_i = \sum_{k\neq i} u_{i,k}^{-2}$, and $\zeta_i = \sum_{k\neq i} u_{i,k}^{-4}$. In the simple case of a unit composed of three transitions, these equations reduce to:
\begin{equation}
	u_1^2 = \frac 1 2 \left(u_{1,2}^2 + u_{1,3}^2 - u_{2,3}^2\right),
\end{equation}
and its permutations, which is reminiscent of the three-cornered hat method~\cite{1536967}.

Finally, the inverse of the squared uncertainties given by equation~(\ref{eq:u2}) yields the weights of each transition in the unit.

\subsection{Truncation of weights}
\label{sec:truncation}

The weights obtained from equations~(\ref{eq:optimalw}) and~(\ref{eq:u2}) are real numbers. However, for the definition to be fully specified, the exact value of these weights must be published. It is therefore practically required to truncate these weights (while keeping their sum equal to one) and to publish them as decimal numbers. This section discusses the consequences of this truncation, considering that because of the truncation, some transitions may be attributed a null weight. For this, two important facts must be highlighted:
\begin{itemize}
    \item There is no practical difference between attributing a negligible weight, and attributing a null weight to a given transition. Indeed, the uncertainty on the recommended frequency $N_i$ barely depends on $w_i$ if $w_i$ is much smaller than the weights of the other transitions composing the unit. This can be seen in equations~(\ref{eq:dN12}): the fractional uncertainty on $N_1$ is proportional to the weight $w_2$ of the other transition, which is close to one if $w_1 \ll w_2$. We can also see such characteristics in equation (\ref{eq:dNk}) for the case of more than three transitions in the pool. Graphically, the projection on the vertical axis of the intersection between the uncertainty area of the frequency ratio and the definition line is unchanged if the definition is vertical or almost vertical. It therefore does not bring any practical advantage to attribute a small weight to some transitions to ensure that they can be used for applications such as  the steering of TAI: they will equally be suitable for these applications, regardless of whether they are assigned a symbolic weight or not.
    \item For all transitions, the uncertainties $u_i$ only have a few significant digits. Because the weights are derived from these uncertainties, they also have a few significant digits. Furthermore, truncating the weights at a few decimal places does not add any uncertainty on the realisation of the unit, because the fixed values of the weights define themselves the unit: ultimately, the choice between the different species of option 1, or option 2, correspond to drastically different choices of weights and they are all currently considered as viable alternatives for the redefinition.
\end{itemize}
From these considerations, it is reasonable to require that the weights are truncated at the second decimal place, in effect excluding transitions with a weight more than a few tens of times lower than the dominant transitions. Thus, the caesium transition would not be part of the new definition under option 2 due to its estimated weight less than 0.01, as discussed in section~\ref{sec:2021}.

As explained in section~\ref{sec:graphical_representation}, the CIPM currently publishes a list of recommended frequencies~\cite{Margolis_2024, pub.1101046402} of various atomic transitions, for the realisation of the meter and the second, besides the Cs transition on which the Primary Frequency Standards (PFS) are based. A subset of these transitions realised with particularly low uncertainty are designated as Secondary Representations of the SI Second (SRSs), indicating that a realisation thereof (a Secondary Frequency Standard, SFS) is suitable for applications requiring an accurate realisation of the second, such as the steering of TAI. Here, we propose how this classification could be adapted under option 2, in light of the considerations about the weight truncation discussed above:
\begin{itemize}
    \item First, the general list of transitions with a recommended frequency for the realisation of the meter would not be affected, as it their realisations span an uncertainty range much larger than the differences at stake between the current definition, option 1, and option 2.
    \item At the other end, the notion of PFS in force in the current definition and in option 1, does not apply to option 2 because the definition is not based on a single transition. As discussed in~\cite{Lodewyck_2024}, the notion of PFS is in fact not fundamental, because actual realisations of the second always deviate to some extend from the ideal definition, and as such, its absence has no practical consequences. In option 2, the PFS would be superseded by frequency standards realising one or several transitions of the ensemble of transitions entering the definition, \emph{i.e.} the transitions with a non-zero weight after truncation. These transitions could be called the ``Defining Transitions'' (DT), and their realisation ``Defining Frequency Standards'' (DFS).
    \item Lastly, the notion of SFS realising SRSs would remain in option 2, with selection criteria identical to those in use for option 1. This list would notably include some of the transitions whose weights have been truncated to zero, including, at least initially, the Cs transition. Given the absence of PFS in option 2, the names ``SRS'' and ``SFS'' sound out of place, and could be replaced with ``Recommended Transitions for the Realization of the Second'' (RTRS), echoing the list of recommended transitions for the realisation of the meter. Their realisation would then be called ``Recommended Frequency Standards'' (RFS).
\end{itemize}
\begin{figure}
    \begin{center}
        \includegraphics[width=0.8\textwidth]{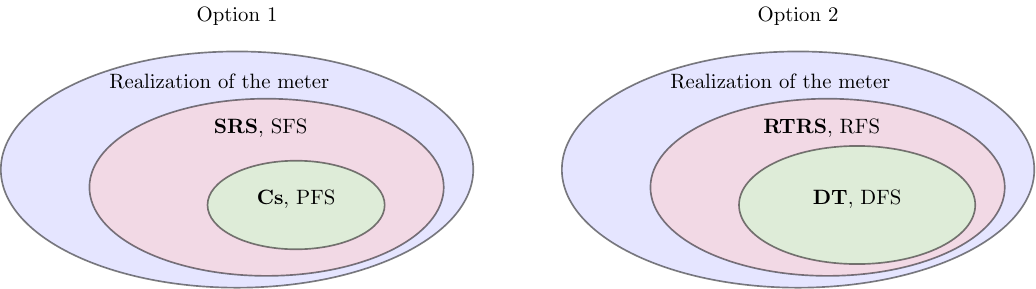}
    \end{center}
    \caption{\label{fig:venn}Hierarchy of frequency standards for the current definition of the SI second or for option 1 after the Cs transition has been replaced by a single optical transition (left); and for option 2 (right). This figures shows three overlapping levels of transitions (bold face) and the associated realisations for the realisation of the meter, the realisation of the second, and the definition of the second. The acronyms are defined in the main text.}
\end{figure}
Figure~\ref{fig:venn} summarises this classification in the form of a Venn diagram.

\subsection{Example using the 2021 adjusted frequency ratios}
\label{sec:2021}

\begin{figure}
	\begin{center}
		\includegraphics[width=0.8\textwidth]{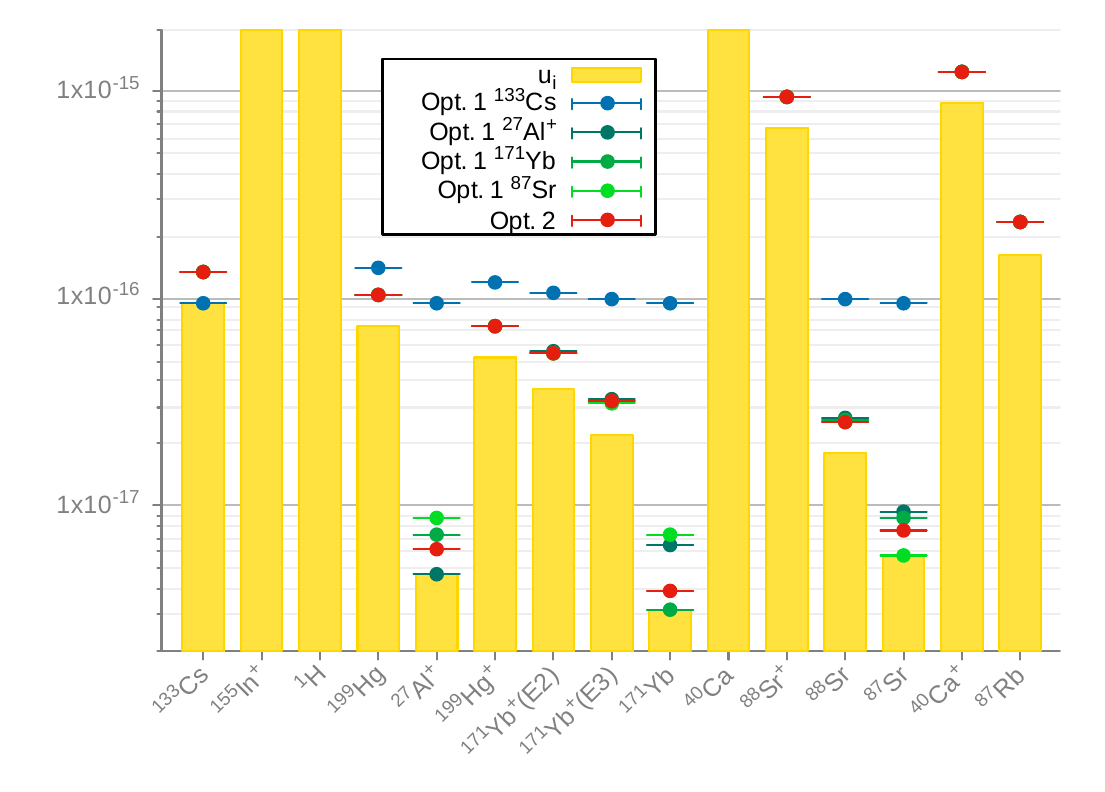}
	\end{center}
	\caption{\label{fig:ui}Graphical representation of the uncertainties presented in table~\ref{tab:2021}, simulating different scenario of a redefinition of the SI second based on the measured frequency ratios considered for the 2021 CCTF. The yellow bar indicate the intrinsic clock uncertainty $u_i$, as estimated using equation~(\ref{eq:u2}). Pointed bars indicate the uncertainty $u_\textrm{tot}$ in realising the unit with a given clock given by equation~(\ref{eq:utot}, \emph{i.e.} the quadratic sum of the clock uncertainty $u_i$ and the uncertainty on the recommended frequency $\delta N_i/N_i$.}
\end{figure}

Using the database of measured frequency ratios used to produce the recommended frequencies of SRSs published for the 2021 CCTF~\cite{fitinput}, we simulate the different options for a redefinition of the SI second, as it would have happened if decided in 2021. The outcome of this procedure is shown in table~\ref{tab:2021}, and plotted in figure~\ref{fig:ui}. For each clock transition, we perform the global fit of frequency ratio as described in the~\ref{sec:globalfit}, in order to calculate the recommended frequencies. As explained in section~\ref{sec:graphical_redefinition}, these recommended frequencies only depend on the set of measured frequency ratios, and on the previous definition of the SI second (here based on ${}^{133}$Cs), but not on the choice of transition(s) entering the new definition. We also report on the typical clock uncertainty $u_i$, calculated using the procedure described in section~\ref{sec:weightsfromu}. The sum of residuals for the optimal uncertainties is $S = 0.57$, much lower than the number of degrees of freedom of the least squares, thus showing a good consistency of the least square optimization.
Three transitions particularly stand out with low uncertainty, namely ${}^{171}$Yb, ${}^{27}$Al$^+$, and ${}^{87}$Sr, as a result of their high precision inter-comparison~\cite{boulder2021frequency}. We then consider five possible options for the redefinition of the SI second: the continuation of the current definition based on caesium, option 1 with either of the three transitions with lowest uncertainty, and option 2 with optimal weights derived from the calculated uncertainties $u_i$. Redefining the second allows to realise the unit with a vastly reduced uncertainty as compared to the current definition (blue points in figure~\ref{fig:ui}). While option 1 allows the realisation of the unit without any overhead beyond the clock uncertainty for the chosen transition (green shaded points aligned with the yellow bars), figure~\ref{fig:ui} shows that option 2 yields, on average, a lower uncertainty (red points). This is already remarkable given that only three transitions significantly contribute to the definition. When more frequency ratios are measured with state-of-the art optical clocks, such as ${}^{171}$Yb$^+$, ${}^{155}$In$^+$, ${}^{88}$Sr$^+$, or ${}^{40}$Ca$^+$, which have already been reported with systematic uncertainties in the $10^{-18}$ range, the red points of option 2 are likely to lower even closer to the yellow bars.

When several clocks that implements different transitions are jointly available, one can combine their output to realise the second. With such a composite system, the uncertainty on the link between the clock ensemble and the definition reads:
\begin{equation}
	\sum_{j,l} (w_j - \delta_j)(w_l - \delta_l)\Sigma'_{jl}
\end{equation}
where $\Sigma'_{jl}$ is the relative correlation matrix between frequency ratios as defined in~\ref{sec:globalfit}, and $\delta_k = 1$ if the transition is realised, or $\delta_k = 0$ otherwise. With the weights indicated in table~\ref{tab:2021}, realising the second with, for instance a ${}^{87}$Sr and a ${}^{171}$Yb would incur an link uncertainty of $1.4\times 10^{-18}$; and down to $4.5\times10^{-19}$ if also using an ${}^{27}$Al$^+$ clock.

\begin{table}
\begin{tabular}{@{}lllllllll}
\br
&&&\centre{4}{Option 1, based on\ldots} & \centre{2}{Option 2}\\
&&&${}^{133}$Cs&${}^{27}$Al$^+$&${}^{171}$Yb&${}^{87}$Sr\\
&&&\crule{4}&\crule{2}\\
Species  & $u_i$ & $N_i$ & $\delta N_i/N_i$ & $\delta N_i/N_i$ & $\delta N_i/N_i$ & $\delta N_i/N_i$ & $w_i$ & $\delta N_i/N_i$\\
 & $\times10^{18}$ &  & $\times10^{18}$ & $\times10^{18}$ & $\times10^{18}$ & $\times10^{18}$ & & $\times10^{18}$\\
\mr
${}^{133}$Cs    & 95 & 9192631770.0000000 & 0 & 96 & 96 & 96 & 0.00 {\color{gray}(0.0006)} & 96\\
${}^{155}$In$^+$   & 2161 & 1267402452901041.283 & 2163 & 2162 & 2162 & 2162 & 0.00 {\color{gray}(0.0000)} & 2162\\
${}^{1}$H       & 3001 & 1233030706593513.654 & 3000 & 3002 & 3002 & 3002 & 0.00 {\color{gray}(0.0000)} & 3002\\
${}^{199}$Hg    & 74 & 1128575290808154.319 & 121 & 74 & 74 & 74 & 0.00 {\color{gray}(0.0010)} & 74\\
${}^{27}$Al$^+$    & 4.7 & 1121015393207859.159 & 96 & 0 & 5.6 & 7.5 & 0.25 {\color{gray}(0.2489)} & 4.1\\
${}^{199}$Hg$^+$   & 53 & 1064721609899146.964 & 109 & 52 & 52 & 53 & 0.00 {\color{gray}(0.0020)} & 52\\
${}^{171}$Yb$^+_\textrm{\tiny E2}$ & 37 & 688358979309308.239 & 102 & 42 & 41 & 41 & 0.00 {\color{gray}(0.0041)} & 41\\
${}^{171}$Yb$^+_\textrm{\tiny E3}$ & 22 & 642121496772645.119 & 97 & 24 & 24 & 23 & 0.01 {\color{gray}(0.0114)} & 23\\
${}^{171}$Yb    & 3.2 & 518295836590863.630 & 96 & 5.6 & 0 & 6.6 & 0.55 {\color{gray}(0.5491)} & 2.2\\
${}^{40}$Ca     & 6276 & 455986240494138.191 & 6276 & 6277 & 6277 & 6277 & 0.00 {\color{gray}(0.0000)} & 6277\\
${}^{88}$Sr$^+$    & 669 & 444779044095486.342 & 667 & 672 & 672 & 672 & 0.00 {\color{gray}(0.0000)} & 672\\
${}^{88}$Sr     & 18 & 429228066418007.006 & 98 & 19 & 19 & 18 & 0.02 {\color{gray}(0.0169)} & 18\\
${}^{87}$Sr     & 5.8 & 429228004229872.992 & 96 & 7.5 & 6.6 & 0 & 0.17 {\color{gray}(0.1658)} & 5.1\\
${}^{40}$Ca$^+$    & 884 & 411042129776400.360 & 885 & 885 & 885 & 885 & 0.00 {\color{gray}(0.0000)} & 885\\
${}^{87}$Rb     & 163 & 6834682610.9043126 & 172 & 166 & 166 & 166 & 0.00 {\color{gray}(0.0002)} & 166\\
\br
\end{tabular}
\caption{\label{tab:2021}Simulation of a redefinition of the SI second based on the frequency ratios measurement published until 2020 that were used as input for the global fit of frequency ratios for the publication of recommended frequencies for the 2021 CCTF. For each clock transition (1st column), we report the adjusted fractional uncertainty calculated using equation~(\ref{eq:u2}) (2nd column), and the recommended frequency (3rd column). The next columns show the uncertainty on the recommended frequency for different possible redefinitions, either based on a single transition (Option 1) or based on the weighted geometric mean of several transitions (Option 2). For the latter, the weights (reported in column 8) are proportional to $1/u_i^2$ (column 2). As proposed in section~\ref{sec:truncation}, the weights are truncated at the second decimal place (for illustration, their value before truncation is reported with 4 significant digits in gray parenthesis). With these truncated weights, the normalisation constant would be $N = 607\,725\,616\,435\,836.99$.  All uncertainties are in units of $10^{-18}$.}
\end{table}

\section{Conclusion}
In this paper, we resolve two main issues that remained about a redefinition of the SI second based on several transitions. The first issue is the difficulty to have an intuitive picture of such a definition, especially about the status of frequency ratios in the definition. We have introduced a graphical representation of the definition in the frequency space in the form of an exactly known curve. With this representation, it is clear that this curve is independent of frequency ratios, and that the latter are only used to ``project'' the exact definition on a given specific transition, \emph{i.e.} to realise the unit with a single specific transition based on a recommended frequency. We have also shown that this graphical representation can be used to illustrate how the recommended frequencies are updated after new frequency ratio measurements, and how the definition curve pivots around the recommended frequencies upon a redefinition of the unit. This graphical representation also bring light to the other options 1 and 3 for the redefinition of the second, but also to the difference between the geometric and arithmetic means for option 2.

The second issue this papers addresses is the determination the weights of the transitions composing the unit. We proposed a rigorous and systematic method that derives the weights from the uncertainties on frequency ratios. The availability of such a methods is indeed a requirement for option 2, and especially for its variation 2b, in which the weights would be regularly updated as the uncertainty of optical frequency standards progresses.

We have thus lifted two major obstacles to the adoption of option 2, enhancing both its general understanding and applicability.

\printbibliography

\clearpage
\appendix

\section{Global fit of frequency ratios}
\label{sec:globalfit}

In this appendix, we recall the notations we use for the global least squares fit of all measured frequency ratios and their correlations, which yields adjusted values $\bar\rho_{i,j}$ for the frequency ratios $\rho_{i,j}$ between the different clock transitions~\cite{pub.1058980983, pub.1101046402}. When considering $n$ clock transitions, there are $n^2$ of these ratios, but only $n-1$ are independent, considering the constraints $\bar\rho_{i,j}\bar\rho_{j,k} = \bar\rho_{i,k}$. Here, we choose these $n-1$ ratios, which are the free parameters of the global fit, to be $\bar\rho_{i,i_0}$, $i \neq i_0$, where  $i_0$ is an arbitrarily chosen clock transition. All the results below are duly independent of this particular choice~\footnote{The coefficients $\Sigma_{j,k}$ of the covariance matrix depend on $i_0$, and are transformed under a change of choice for $i_0$ according to equation~(13) of reference~\cite{pub.1058980983}, under which the uncertainty of any function of the frequency ratios given by equation~(\ref{eq:df}) is invariant. Explicitly, the coefficients of the covariance matrix transform as $\Sigma'^{(i_1)}_{j,k} = \Sigma'^{(i_0)}_{j,k} + \Sigma'^{(i_0)}_{i_1,i_1} - \Sigma'^{(i_0)}_{j,i_1} -\Sigma'^{(i_0)}_{i_1,k} $, when choosing the fixed transition $i_1$ instead of $i_0$.}. Besides yielding adjusted values of $\bar\rho_{i,j}$, the global fit also outputs the covariance matrix $\Sigma$:
\begin{equation}
	\Sigma_{j,k} = 2\left(\frac{\partial^2 \chi^2}{\partial \rho_{j,i_0}\partial \rho_{k,i_0}}\right)^{-1}
\end{equation}
where $\chi^2$ is the sum of the squared residual of the global fit of measured frequency ratios, as defined by equation~(A1) of reference~\cite{lodewyck2019definition}. The matrix $\Sigma$ is a $n-1$ by $n-1$ symmetric matrix, but we can extend it to a $n$ by $n$ matrix by inserting a zero row and a zero column at index $i_0$, thus facilitation the indexing of clock transitions. The covariance matrix $\Sigma$ can also be expressed in relative units:
\begin{equation}
	\Sigma'_{j,k} \equiv \frac{\Sigma_{j,k}}{\bar\rho_{j,i_0} \bar\rho_{k,i_0}}
\end{equation}
The uncertainty on a function $f$ of the frequency ratios can be deduced from the covariance matrix with the formula:
\begin{equation}
	\label{eq:df}
	(\delta f)^2 = \sum_{j,k} \frac{\partial f}{\partial \rho_{j,i_0}}\frac{\partial f}{\partial \rho_{k,i_0}} \Sigma_{j,k}
\end{equation}
For instance, we have:
\begin{equation}
	\label{eq:deltarhoij}
	\left(\frac{\delta \bar \rho_{i,j}}{\bar \rho_{i,j}}\right)^2 = \Sigma'_{ii} + \Sigma'_{jj} - 2 \Sigma'_{ij},
\end{equation}
a result which holds for $i = i_0$ and/or $j = i_0$ considering the extension of $\Sigma$ to a $n\times n$ matrix. Applying formula~(\ref{eq:df}) to the expression of recommended frequencies of option 2:
\begin{equation}
	N_k = N \prod_m \rho_{k,m}^{w_m} = N \rho_{k,i_0} \prod_m \rho_{i_0,m}^{w_m}, \quad \textrm{hence}\quad \frac{\partial N_k}{\partial \rho_{j,i_0}} = \frac{\delta_{jk}-w_j}{\rho_{j,i_0}}N_k,
\end{equation}
we can express the relative uncertainty on the recommended frequency of the transition $k$ by:
\begin{equation}
	\label{eq:dNk}
	\left(\frac{\delta N_k}{N_k}\right)^2 = \sum_{j,l} (w_j - \delta_{jk})(w_l - \delta_{lk})\Sigma'_{jl}
\end{equation}

\section{Proof of equation~(\ref{eq:optimalw})}
\label{sec:proofoptimal}

Using equation~(\ref{eq:dNk}), we can write equation~(\ref{eq:leastsquaresw}) as:
\begin{eqnarray}
	&& \frac{\partial R}{\partial w_i} = 0 \\
	&\Leftrightarrow& \sum_{k,j,l} \frac{\Sigma'_{jl}}{u_k^2} \frac{\partial}{\partial w_i}\left((w_j - \delta_{jk})(w_l - \delta_{lk})\right) = 0 \\
	&\Leftrightarrow& \sum_{k,j,l} \frac{\Sigma'_{jl}}{u_k^2}\left(\delta_{ij}(w_l - \delta_{lk})+ \delta_{il}(w_j - \delta_{jk})\right) = 0 \\
	&\Leftrightarrow&
	\sum_{k,j,l} \frac{\Sigma'_{jl}}{u_k^2}\delta_{ij}w_l +
	\sum_{k,j,l} \frac{\Sigma'_{jl}}{u_k^2}\delta_{il}w_j =
	\sum_{k,j,l} \frac{\Sigma'_{jl}}{u_k^2}\delta_{ij}\delta_{lk} +
	\sum_{k,j,l} \frac{\Sigma'_{jl}}{u_k^2}\delta_{il}\delta_{jk}
	 \\
	&\Leftrightarrow&
	\sum_{l} \Sigma'_{il}w_l = \frac{1}{\sum_k \frac{1}{u_k^2}}\left(
	\sum_{l} \frac{\Sigma'_{il}}{u_l^2}\right)
\end{eqnarray}
The solution for this set of equations is:
\begin{equation}
	w_l = \frac{\frac{1}{u_l^2}}{\sum_k \frac{1}{u_k^2}},
\end{equation}

\section{Graphical representation of option 3}
\label{sec:graphicaloption3}

\begin{figure}
	\begin{center}
		\includegraphics[width=0.8\textwidth]{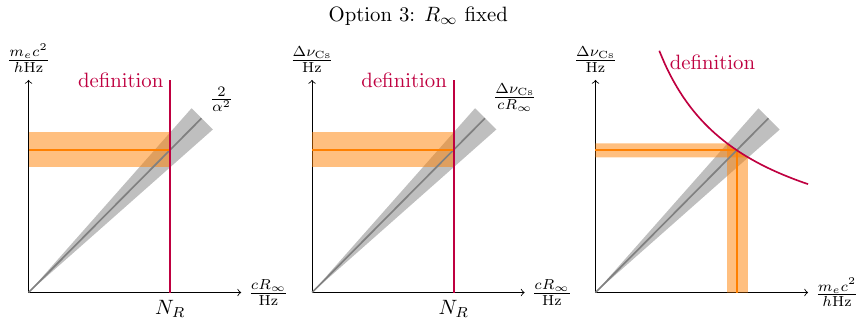}
	\end{center}
	\caption{\label{fig:graphicaloption3}
	Graphical representation of the definition for option 3, in which the unit of time is set by fixing the numerical vale of the Rydberg constant to $N_R$~Hz exactly. The left graphs shows the definition in the (electron mass, Rydberg constant) plane, the middle graph in the ($\Dnucs$, Rydberg constant plane), and the right graph in the ($\Dnucs$, electron mass plane). These graphs show a strong similarity with the graphs of options 1 and 2, represented in figure~\ref{fig:graphical}. Note that on the right graph, the definition should be represented with an uncertainty area accounting for the fact the function $f$ linking the axis also depends on not perfectly known quantities. This artifact vanishes when considering a larger dimensional space including all the quantities entering $\Dnucs/\textrm{Hz}$, such as $m_p/$kg, \ldots
}
\end{figure}

In addition to options 1 and 2 described above, the roadmap for the redefinition of the SI second~\cite{10.1088/1681-7575/ad17d2} also mentions option 3, which consists in defining the second by fixing a fundamental constant. Although conceptually appealing, and in line with the policy of the redefinition of the SI units in 2019, this option is not practical for the case of the redefinition of the second, because of the large uncertainty on these constant. In this appendix, we nevertheless imagine a system of units following option 3, in which the Rydberg constant is attributed a fixed, exact, constant $N_R$ when expressed in Hz:
\begin{equation}
	c R_\infty \equiv N_R \textrm{ Hz}
\end{equation}
while the numerical values of $c$ and $h$ are also kept constant, to $N_c$~m/s and $N_h$~J$\cdot$s respectively, as they are in the current SI. The relation between the Rydberg constant and the electron mass $m_e$ reads:
\begin{equation}
	\label{eq:rydberg}
	c R_\infty = \frac{m_e c^2}{2h} \alpha^2,
\end{equation}
from which we can write the numerical value of the electron mass $m_e$, or equivalently its Compton frequency $\nu_e = m_e c^2/h$, as:
\begin{equation}
	\label{eq:me}
	\frac{m_e}{\textrm{kg}} = \frac{2 N_h N_R}{N_c^2} \frac{1}{\alpha^2} \qquad \textrm{and} \quad \frac{\nu_e}{\textrm{Hz}} = 2 N_R \frac{1}{\alpha^2}
\end{equation}
In this equation, the numbers $N_h$, $N_R$, and $N_c$ have exactly known but arbitrary, human-chosen values~; on the contrary, the fine structure constant $\alpha$ is a dimension-less number set by nature, and is only accessible to us by experimental measurements, which have a limited uncertainty. As a consequence, the numerical value of $m_e$ in our system of units is fundamentally limited by our knowledge of $\alpha$. These considerations can be represented graphically, following the same method as developed in this paper (see section~\ref{sec:graphical}): the left graph of figure~\ref{fig:graphicaloption3} illustrates that the numerical value of the electron Compton frequency is at the intersection of the fixed numerical value of the Rydberg constant and the line of slope $\alpha^2/2$. This is analogous to Figure~\ref{fig:graphical}~(a), where the role of the frequency ratio $\rho_{2,1}$ is played by $\alpha$.

In the system of units under consideration here, the frequency in Hz of the caesium clock transition is not fixed to a known value, but has an uncertainty, arising from the experimental determination of the frequency ratio $\Dnucs/cR_\infty$, obtained for instance by comparing a caesium clock with an hydrogen spectroscopy setup. This uncertainty is represented in the middle graph of figure~\ref{fig:graphicaloption3} in the plane of coordinates $\Dnucs/\textrm{Hz}$ and $cR_\infty/\textrm{Hz}$.

We may now wonder how to graphically represent the definition in the plane whose coordinates
are the numerical values of the caesium clock frequency and of the electron Compton frequency. Because none of these two quantities constitutes the definition of the unit, we have to apply the laws of physics to write:
\begin{equation}
	\label{eq:dnucsf}
	\Dnucs = cR_\infty f(\alpha, \frac{m_e}{m_p}, \Lambda_\textrm{\tiny QCD}, \ldots)
\end{equation}
where the function $f$ of dimension-less arguments gathers all the complex physics of the interaction of the Cs many electrons with its nucleus, and with the zero point energy of the surrounding quantum fields. Although the function $f$ is desperately hard the compute, its expression is, in principle, accessible without any experimental uncertainty by correctly solving the laws of physics\footnote{As opposed to the values of its arguments, which are, however, subject to experimental uncertainty.}. Equation~(\ref{eq:dnucsf}) can be rewritten as:
\begin{equation}
	\frac{\Dnucs}{\textrm{Hz}} = N_R f(\alpha, \frac{m_e}{m_p}, \Lambda_\textrm{\tiny QCD}, \ldots),
\end{equation}
exhibiting one of our plane coordinates $\Dnucs/\textrm{Hz}$. Where does the second coordinate $\nu_e/\textrm{Hz}$ appear? Although $m_e$ appears in the second argument of $f$, its numerical value in the system of unit does not play a role in this argument, because $m_e/m_p$ is a dimension-less quantity independent of the numerical value of $\nu_e$ in the system of unit. However, according to equation~(\ref{eq:me}), $\alpha$ is a direct function of the numerical value of $\nu_e$ in the system of units, such that
\begin{equation}
	\label{eq:dnucsfue}
	\frac{\Dnucs}{\textrm{Hz}} = N_R f\left(\left(\frac{1}{2N_R} \frac{\nu_e}{\textrm{Hz}}\right)^{-\frac 1 2}, \frac{m_e}{m_p}, \Lambda_\textrm{\tiny QCD}, \ldots\right),
\end{equation}
As $f$ can be approximated by a power law function, the curve representing equation~(\ref{eq:dnucsfue}), as depicted in the right graph of figure~\ref{fig:graphicaloption3}) is similar to the curved representation of the definition of option 2, depicted  in the second graph of figure~\ref{fig:graphical}. This shows that the graphical representation proposed in this paper provides a consolidated picture not only for option 1 and 2, but also for option 3. ; and that the notion of geometrical mean also appears in the more fundamental nature of option 3.

\section{Geometric mean vs arithmetic mean: a graphical point of view}
\label{sec:artihmvsgeom}

\begin{figure}
	\begin{center}
		\includegraphics[width=0.3\textwidth]{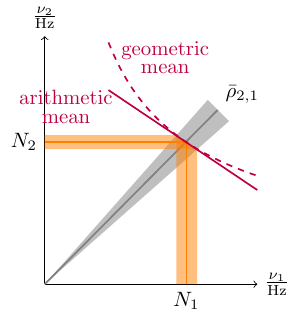}
	\end{center}
	\caption{\label{fig:graphical_arithmetic}Graphical representation of the arithmetic mean by a straight line (solid purple line) in the $(\nu_1/\textrm{Hz}, \nu_2/\textrm{Hz})$ plane. The arithmetic mean is the tangent to the geometric mean representation (dashed line). Although both means are practically equivalent, the geometric mean is conceptually more adapted as a multi-species definition of the second, and yields simpler formulas.}
\end{figure}

Instead of defining the SI second by fixing the weighted geometric mean of several clock frequencies, one could rather think of using the weighted arithmetic mean. Such a definition would be written:
\begin{equation}
    \label{eq:arithmetic}
    1~\textrm{Hz} \equiv \sum_{k \in \mathcal{C}} a_k \nu_k.
\end{equation}
The frequency of the transition $i$ therefore reads:
\begin{equation}
    \nu_i = \frac{1}{\sum_{k \in \mathcal{C}} a_k \rho_{k,i}}\,\textrm{Hz},
\end{equation}
from which we can define the numerical values $N_i$ of the recommended frequencies as the best estimate of $\nu_i/$Hz:
\begin{equation}
    N_i = \frac{1}{\sum_{k \in \mathcal{C}} a_k \bar\rho_{k,i}},
\end{equation}
using the output $\bar\rho_{k,i}$ of the global fit of frequency ratios. Finally, by applying the relation~(\ref{eq:df}), we can calculate the uncertainty of these recommended frequencies as:
\begin{equation}
	\label{eq:dNkarithm}
    \left(\frac{\delta N_k}{N_k}\right)^2 = \sum_{j,l} (N_ja_j - \delta_{jk})(N_la_l - \delta_{lk})\Sigma'_{jl}
\end{equation}

In~\cite{lodewyck2019definition}, it was argued that such a definition has conceptual shortcomings, and that replacing the arithmetic mean with the geometric mean is a \emph{sufficient} condition to resolve them. In this appendix, we take advantage of the graphical representation introduced in section~\ref{sec:graphical} to give a more intuitive explanation of these shortcomings, and to show that the geometric mean is actually a \emph{necessary} condition to resolve them.

Figure~\ref{fig:graphical_arithmetic} provides a graphical representation of the arithmetic mean~(\ref{eq:arithmetic}) for the special case of two transitions. With such a mean, the definition curve is:
\begin{equation}
    \label{eq:arithmeticcurve}
	a_1 \frac{\nu_1}{\textrm{Hz}} + a_2 \frac{\nu_2}{\textrm{Hz}} = 1.
\end{equation}
Unlike the geometric mean, it is represented by a straight line in the $(\nu_1/\textrm{Hz}, \nu_2/\textrm{Hz})$ plane, albeit not a vertical or horizontal line as for option 1. The slope of this line is $-a_1/a_2$. As described in section~\ref{sec:graphical}, the numerical values $N_1$ and $N_2$ of the  recommended frequencies are at the intersection between the definition curve and the line of slope $\bar\rho_{2,1}$, such that
\begin{equation}
    \label{eq:relN1N2arithm}
    a_1 N_1 + a_2N_2 = 1, \qquad \textrm{and} \qquad \frac{N_2}{N_1} = \bar\rho_{2,1},
\end{equation}
and the uncertainties on the recommended frequencies are the projections of the intersection between the definition curve and the uncertainty on the frequency ratio. Following the same calculation as in section~\ref{sec:graphical_representation}, or specializing equation~(\ref{eq:dNkarithm}) for two transitions, gives the relations:
\begin{equation}
    \label{eq:dN12arithm}
    \frac{\delta N_1}{N_1} = a_2N_2\frac{\delta \bar\rho_{2,1}}{\bar\rho_{2,1}}, \qquad \frac{\delta N_2}{N_2} = a_1 N_1\frac{\delta \bar\rho_{2,1}}{\bar\rho_{2,1}},
\end{equation}
which are arguably less elegant than their counterparts~(\ref{eq:dN12}) for the geometric mean, because the relation between the fractional uncertainty on the recommended frequencies and the fractional uncertainty on the frequency ratio depends on the recommended frequencies.

We can nevertheless apply the procedure described in section~\ref{sec:weightsfromu} in order to relate the coefficients $a_i$ to the clock uncertainties $u_i$. Comparing equation~(\ref{eq:dNkarithm}) to equation~(\ref{eq:dNk}), (or for two transitions, comparing equation~(\ref{eq:dN12arithm}) to equation~(\ref{eq:dN12})), it is clear that this procedure yields the optimal coefficients:
\begin{equation}
    \label{eq:optai}
	a_i = N_iw_i,
\end{equation}
where $w_i$ is defined by equation~(\ref{eq:optimalw}). This relation was already introduced in~\cite{lodewyck2019definition}, and intuitively states that the arithmetic weights $a_i$ should be scaled by the clock frequencies, to avoid that a transition has a larger weight just by virtue of its larger frequency. In equation~(\ref{eq:optai}), the quantities $w_i$ thus correspond to the "physical weights" that should be attributed to each transition, solely related to their respective realisation uncertainties, and are identical to the weights of the geometric mean. Injecting~(\ref{eq:optai}) into~(\ref{eq:dN12arithm}), we fall back on~(\ref{eq:dN12}): this shows that the definition curve of the arithmetic mean is the tangent to the definition curve of the geometric mean at point $(N_1, N_2)$, as expected.

Now, let's consider that new frequency ratio measurements result in a change of recommended frequency ratio $\bar\rho_{2,1}$. After this change, the coefficients $a_1$ and $a_2$ should stay constant, because the definition of the unit should not be altered by new frequency ratio measurements, as discussed in section~\ref{sec:graphical_updateratios}. But we have, from equations ~(\ref{eq:relN1N2arithm}) and~(\ref{eq:optai}):
\begin{equation}
	\label{eq:fracratios}
	\frac{a_1}{a_2} = \bar \rho_{2,1} \frac{w_1}{w_2},
\end{equation}
This means that the physical weights $w_i$ must change, which leads to a contradiction if, for instance, the new frequency ratio measurements are performed without improving the uncertainties of the clocks. This contradiction could have been anticipated by initially choosing a definition with a steeper slope for larger $\bar \rho_{2,1}$, and a shallower slope for smaller $\bar \rho_{2,1}$, so that equation~(\ref{eq:fracratios}) always holds in the vicinity of $(N_1, N_2)$. In other words, we should have chosen a curved definition, instead of the straight line~(\ref{eq:arithmeticcurve}). Noting $f(x)$ the function represented by this curve, \emph{i.e} such that $\nu_2/\textrm{Hz} = f(\nu_1/\textrm{Hz})$, equation~(\ref{eq:fracratios}) can be rewritten as:
\begin{equation}
	-f'(x) =  \frac{f(x)}{x} \frac{w_1}{w_2},
\end{equation}
whose solution is $f(x) = N^{\frac{1}{w_2}} x^{-\frac{w_1}{w_2}}$, where $N^{\frac{1}{w_2}}$ is an integration constant. Therefore, we have $\nu_1^{w_1}\nu_2^{w_2} = N\,\textrm{Hz}$: this is the geometric mean.

In conclusion, the geometric mean can be viewed as the function that preserves the physical weights of the transitions under a change of estimated frequency ratios. We note however that this discussion is only conceptual: in practice, the variations of the physical weights occurring with the arithmetic mean, on the same order as the relative variations of frequency ratios, are many orders of magnitude smaller than the truncation of the weights discussed in section~\ref{sec:truncation}.

\end{document}